\documentclass[reprint,amsmath,amssymb,aps,nofootinbib]{revtex4-1}
\usepackage{hyperref}
\usepackage[utf8]{inputenc}
\usepackage{textcomp}
\usepackage{graphicx}
\usepackage{t1enc}
\usepackage{amsmath}
\usepackage{mathrsfs}
\usepackage[T1]{fontenc}
\usepackage{amsfonts}
\usepackage{color}
\usepackage{wasysym}

\newcommand{\ud}{\mathrm{d}}
\newcommand{\ui}{\mathrm{i}}

\begin{document}
\title{Model for Quantum Effects in Stellar Collapse}
\author{Bruno Arderucio-Costa}
\email{arderucio@physics.ubc.ca}
\author{William G. Unruh}
\email{unruh@physics.ubc.ca}
\affiliation{Dept. of Physics and Astronomy, University of British Columbia}

\date{\today}

\begin{abstract}
We present a simple model for stellar collapse and evaluate the quantum mechanical stress-energy tensor to argue that quantum effects do not play an important role for the collapse of astrophysical objects.
\end{abstract}

\maketitle

\section{Introduction}
Over the course of the last few years, there has been proposed a number of models to prevent information loss in a black hole. Most popular versions propose small modifications to either General Relativity or Quantum Mechanics. But it has been argued \cite{LauraB,Laura, refB, Terno, ho} that black holes could simply not form as a result of the back reaction of its Hawking emission. While it is well known that all outgoing null rays that carry low energy at infinity pile up at the horizon, we show that this fact does not mean that one can find an infinitely large amount energy on the horizon itself as a result of the non conservation of the expectation value of the outgoing energy flux along these lines.

In section 2 we create a 2-dimensional model for a collapsing star based on the Oppenheimer-Snyder model (rather than the collapsing null shell in \cite{DFU}) resulting in a black hole and in section 3 we compute the expectation values of the regularized stress-energy tensor of a massless scalar filed in that spacetime and we conclude in section 4.

\section{Model for Stellar Collapse}
We model our 4-dimensional collapsing star as dust (zero pressure) as a FRW solution\footnote{The form of FRW solution below, referred as $k=1$ is necessary so Einstein's equations are compatible with initial condition of the dust being at rest in the comoving frame.} in the interior and Schwarzschild in the exterior of the star.
\begin{widetext}
\begin{equation}
 \ud s^2=
 \begin{cases}
  a^2(\eta)(-\ud\eta^2+\ud\chi^2+\sin^2\chi\ \ud\Omega^2) & \text{interior}\\
  -\left(1-\frac{2m}{r}\right)\ud t^2+\left(1-\frac{2m}{r}\right)^{-1}\ud r^2+r^2\ud\Omega^2 & \text{exterior},
  \end{cases}
 \label{metric}
\end{equation}
\end{widetext}
where $a(\eta)=\frac{a_0}{2}(1+\cos\eta),\ 0\leq\eta\leq\pi$.

The surface of the star is described in the inner region by $\chi=\chi_0$ and in the exterior is described by a radial timelike geodesic $R(t)$ passing through the point $r=a_0\sin\chi_0\equiv R_0$. Exterior region's geodesic equations for stellar surface leads to
\begin{equation}
 \frac{\ud t}{\ud R}=\sqrt{\frac{1-2m/R_0}{2m/R-2m/R_0}}\left(1-\frac{2m}{R}\right)^{-1}.
 \label{SSurface}
\end{equation}

Israel junction conditions for non singular energy-momentum tensor on the interface (that is, continuity of the induced metric and the extrinsic curvature across the junction interface) force \cite{George, rc1}
\begin{equation}
 m=\frac{a_0}{2}\sin^3\chi_0
 \label{junction1}
\end{equation}
and
\begin{equation}
 \eta=\arccos\left(\frac{2R(t)}{a_0\sin\chi_0}-1\right)
 \label{junction2}
\end{equation}
 on the interface of the two regions.

The event horizon is located by finding the outgoing null geodesic that coincides with $r=2m$ in the exterior region to give 
\begin{equation}
 \eta=\chi-\chi_0+\arccos\left(\frac{4m}{R_0}-1\right).
\label{eventhorizon}
\end{equation}

\section{Expectation Value of Regularized Stress-Energy Tensor}

\subsection{2-dimensional case}
Now we consider the section $\ud\Omega=0$ of spacetime manifold so we can cast the inner metric in a conformally flat form $\ud s^2=-a^2(U,V)\ud U\ud V$, $a(U,V)=\frac{a_0}{2}\left(1+\cos\frac{U+V}{2}\right)$. For later convenience we choose the origin of our advanced and retarded coordinates so that $U=\eta-\chi+\chi_0$ and $V=\eta+\chi-\chi_0$ so that the surface of the collapsing body is simply $U=V$.
The curvature scalar can be readily computed from the 2D resulting metric to give

\begin{equation}
 \mathscr R=
 \begin{cases}
  \frac{2}{a^4}(aa''-a'^2)= \frac{8}{a_0^2(1+\cos\eta)^3} & \text{interior} \\
  \frac{4m}{r^3}=\frac{2a_0\sin^3\chi_0}{r^3}, & \text{exterior}
 \end{cases}
 \label{ricciscalar}
\end{equation}
where prime denotes differentiation with respect to $\eta$.

One can employ the prescription of \cite{DFU} to compute the expectation value of the energy momentum tensor after covariant point splitting regularization with respect to the vacuum state defined by coordinates $\bar u$ and $\bar v$, using the same notation of that reference.

\subsection{Choice of Vacuum State}
Our modes $\bar u$ and $\bar v$ are defined with respect to $\mathscr I^-$ to mimic the so-called Unruh vacuum so $\bar v=t+r+2m\log(r-2m)$ is the usual Schwarzschild advanced coordinate. An ingoing null geodesic coming from $\mathscr I^-$ passes through the surface of the collapsing object and a mode proportional to $e^{\ui\omega\bar v}$ at $\mathscr I^-$ will be reflected at $\chi=0\iff V=U-2\chi_0=U-2\arcsin\sqrt{\frac{2m}{R_0}}$ (using (\ref{junction1}) and (\ref{junction2}) in the last step) and come out to $\mathscr I^+$ like $e^{\ui\omega\bar u}$ with 
\begin{equation}
\bar u=v\left(U(u)-2\arcsin\sqrt{\frac{2m}{R_0}}\right)
\label{ubar}
\end{equation}
at the interface, where the composition $U(V)$ is evaluated at the surface of the collapsing object using (\ref{junction2}) to relate these coordinates on the interface:
\begin{equation}
U(u)=\arccos\left(\frac{2R}{R_0}-1\right)\Rightarrow U'(u)=\frac{R-2m}{2R\sqrt{R(R_0-R)}}
 \label{alpha}
\end{equation}
and
\begin{multline}
v(V)=t\left(\frac{R_0}{2}(1+\cos V)\right)+R^*\left(\frac{R_0}{2}(1+\cos V)\right)-R_0^*\Rightarrow \\
v'(V)=\frac{R_0^2\sin V\cos^2\frac{V}{2}}{2[R_0(1+\cos V)-4m]}\left(\sqrt{\frac{2(R_0-2m)\cot^2\frac{V}{2}}{m}}-2\right),
 \label{beta}
\end{multline}
where (\ref{SSurface}) was used for taking derivative of the first term in (\ref{beta}) and $R^*(r)\equiv r+2m\log(r-2m)$ and $R_0^*$ is an abbreviation for $R^*(R_0)$.

There are no contributions to the expectation value of the energy-momentum tensor coming from the maching conditions of the field modes across the collapsing surface. Because the surface where the matching conditions are applied is a geodesic and satisfy Israel junction conditions with no $\delta$ terms in its energy-momentum tensor, it is required that the field modes and its derivatives across the surface are continuous\cite{rc2}. These conditions are satisfied, since the solutions to the field equations are merely the composition between the solution in one chart and the matching conditions (\ref{alpha}) and (\ref{beta}), both $\mathscr C^1$ functions in their variables.

\subsection{Results}
After covariant point-splitting regularization, the components of the expectation value of the stress-energy tensor for the 2D spacetime is given by \cite{DFU}
\begin{equation}
 \langle T_{\mu\nu}\rangle=\frac{\mathscr R}{48\pi}g_{\mu\nu}+\theta_{\mu\nu},
 \label{VEV}
\end{equation}
with $\theta_{\mu\nu}$ defined in that reference.
Introducing the abbreviation $$\mathcal F(f(x);x)\equiv\frac{\sqrt{f}}{12\pi}\frac{\ud^2}{\ud x^2}\frac{1}{\sqrt{f}}=\frac{1}{24\pi f}\left(\frac{3(\ud f/\ud x)^2}{2f}-\frac{\ud^2f}{\ud x^2}\right),$$
\[\mathcal F(fg;x)=g\mathcal F(f;x)+f\mathcal F(g;x)-\frac{f'g'}{24\pi fg},\]
the components of expectation value of the renormalized stress-energy tensor are given by \cite{DFU}
\begin{gather}
 \langle0|T_{vv}|0\rangle=\frac{m(3m-2r)}{48\pi r^4}\\
 \langle0|T_{uv}|0\rangle=\frac{m(2m-r)}{24\pi r^4}\\
 \langle0|T_{uu}|0\rangle=\left(\frac{\ud\bar u}{\ud u}\right)^2\mathcal F\left(\left(\frac{\ud u}{\ud \bar u}\right)\left(1-\frac{2m}{r}\right);\bar u\right).
 \label{Text}
\end{gather}
outside the collapsing body and \cite{davies, BirrelDavies}
\begin{gather}
 \langle0|T_{UU}|0\rangle=\mathcal F(v'(U-2\chi_0);U)-\mathcal F(a(U);U)\\
 \langle0|T_{VV}|0\rangle=\mathcal F(v^\prime(V);V)-\mathcal F(a(V);V)
 \label{Tint}
\end{gather}
inside it.

\subsubsection{Inside Matter}

Explicit evaluation gives
\[\mathcal F(a(U);U)=\frac{\csc^2\frac{U}{2}(3+\cos U)}{96\pi},\]
\begin{widetext}
\begin{multline*}
\mathcal F(v^\prime(V);V)=\frac{1}{192\pi m (R_0+R_0\cos V-4m)^2(\xi-2)^2}\times\\
\left\{ m^3(256(1-4\cos 2V)-\xi(1792-512\cos 2V)+256\cos V\csc^2V(9\xi-10))\right.+\\
\left.m^2R_0(1088+\xi(224+96\cos3V)-4\cos2V(48+56\xi)-32\cos V\csc^2V(3\xi-34))\right.+\\
\left.mR_0^2(56\cos3V-412+2\cos4V(2-\xi)-6\xi-16\cos2V(1-\xi)-472\cos V\csc^2V)\right.+\\
\left.R_0^3(45-\cos4V-20\cos2V+64\cos V\csc^2V)\right\},
\end{multline*}
\end{widetext}
where $\xi\equiv\sqrt{\frac{2(R_0-2m)\cot^2\frac{V}{2}}{m}}$. The missing term $\mathcal F(v^\prime(U-2\chi_0);U)$ is obtained from above by replacing $V$ by $U-2\chi_0$ in $\mathcal F(v^\prime(V);V)$ and $\mathcal F(a(V);V)$ by replacing $U$ by $V$ in  $\mathcal F(a(U);U)$.

The outgoing energy flux as measured by a stationary observer within the star $\langle0|T_{UU}|0\rangle$ on the event horizon $U=\arccos(4m/R_0-1)$ is finite for all $\chi_0\neq0$, including in its formation, and all the way until the horizon intersects the surface of the star, in explicit disagreement with \cite{Laura}.

\subsubsection{Outside Matter}
The formulas for $\langle0|T_{vv}|0\rangle$ and $\langle0|T_{uv}|0\rangle$ are the same as in \cite{DFU}, whilst the explicit formula for $\langle0|T_{uu}|0\rangle$ is several pages long and it is more elucidative to show a plot (figure 1) than writing it. There are some properties that are worthwhile elaborating. First, we note that at the onset of the collapse from (\ref{ubar}) $\bar u=u$, so that  $\langle0|T_{uu}|0\rangle=\langle0|T_{vv}|0\rangle$, while at very late times ($R\to2m+0$) it restores Hawking's result $\kappa^2/4\pi=1/(768\pi m^2)$ as $r\to\infty$ for any initial radius, as one would expect. More generally, for very late times, $\langle0|T_{uu}|0\rangle$ as a function of $r$ turns out to be the same as in \cite{DFU}, suggesting that the nature of the collapse does not affect the final energy-momentum tensor of the radiation emitted by the black hole. The only divergence present in this expression is for the true singularity at the end of the stellar collapse at $r=0$.

\begin{figure}[h]
\begin{center}
 \includegraphics[width=\columnwidth]{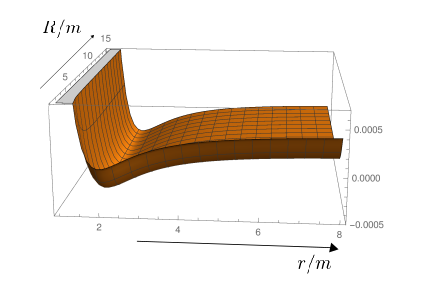}
 \end{center}
 \caption{$m^2\langle0|T_{uu}|0\rangle$ as a function of $r$ coincides with $\langle0|T_{vv}|0\rangle$ on the onset of the collapse, here represented by large values of $R$ and with \cite{DFU} for very late times, for $R\to2m$. The horizontal plane on top of the graph represents the constant $\kappa^2/4\pi$.}
\end{figure}

To study possible effects of back-reaction, it is interesting to analyze observables measured by an observer following the surface of the collapsing star extracted from the results above, namely $\langle T_{ab}\rangle\ell^a\ell^b$, $\langle T_{ab}\rangle\ell^am^b$ and $\langle T_{ab}\rangle m^am^b$ where $(\ell^a,m^a)$ is a diad adapted to this observer, so $\ell^a$ is the tangent vector to the ingoing timelike geodesic describing the stellar surface, and $m^a$ is a spacelike unit vector orthogonal to $\ell^a$. Using the Eddington-Finkelstein coordinate system\footnote{These coordinates are more adapted to study the neighbourhood of the future horizon.} to write these vectors,
\begin{figure}[h]
 \begin{center}
  \includegraphics[width=\columnwidth]{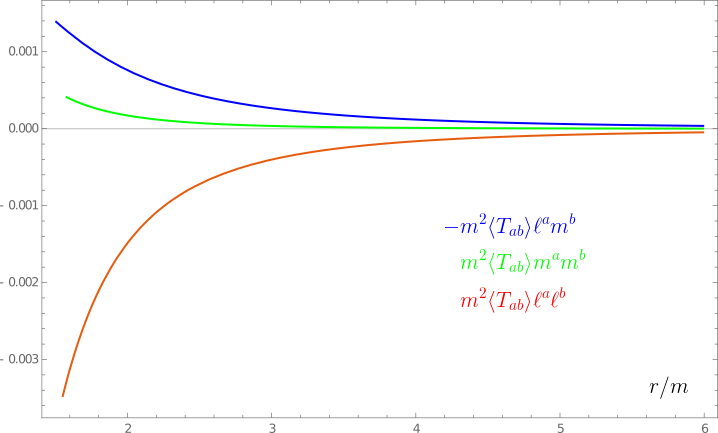}
 \end{center}
\caption{All appropriate observables as measured by someone following the collapse obtained from the regularized energy-momentum tensor are finite everywhere except at the singularity. The graph shows exclusively the region around $r=2m$.}
\end{figure}

\begin{widetext}
\begin{equation}
\ell^a=\left(1-\frac{2m}{r}\right)^{-1}\left(\sqrt{1-\frac{2m}{R_0}}-\sqrt{\frac{2m}{r}-\frac{2m}{R_0}}\right)\left(\frac{\partial}{\partial v}\right)^a-\sqrt{\frac{2m}{r}-\frac{2m}{R_0}}\left(\frac{\partial}{\partial r}\right)^a,
\label{tanvec}
\end{equation}

and

\begin{multline}
 -m^a=\sqrt{\frac{r\left(R_0r+2mR_0-4mr-2rR_0\sqrt{(2m/r-2m/R_0)(1-2m/R_0)}\right)}{R_0(r-2m)^2}}\left(\frac{\partial}{\partial v}\right)^a+\\
 \frac{R_0\sqrt{(2m/r-2m/R_0)(1-2m/R_0)}+R_0-2m}{R_0}m^v\left(\frac{\partial}{\partial r}\right)^a.
 \label{ortvec}
\end{multline}
\end{widetext}

All three contractions above are regular everywhere except at the singularity, as shown in figure 2.

For an observer in the interior of the matter, all these observables are also finite, as one can see from the results from section 3.3.1 and by realizing that the vector components of the vector tangent to timelike geodesics in coordinates (\ref{metric}) are regular, and so are the components of the spacelike vector orthogonal to it, since the metric components in these coordinates are regular and invertible.

For observers outside the the star, e.g. for a geodesic observer starting from infinity, observables can be found by the same method. If the observer starts at rest at infinity, the tangent vector and the spacelike vector orthogonal to it can be found from (\ref{tanvec}) and (\ref{ortvec}) respectively by making $R_0\to\infty$. 

\section{Discussion}
From (\ref{VEV}), (\ref{ricciscalar}) and (\ref{Text}-\ref{Tint}) we see that, in contrast to \cite{Laura}, the regularized energy-momentum tensor is perfectly regular at the classical event horizon and small for astrophysical black holes, ($\sim 1/768\pi m^{2}$ for $r=2m$)
the only divergence occurs at the singularity when the star collapse entirely at $\eta=\pi$, where the classical stress-energy tensor is also divergent as the scale parameter goes to zero. The same is true for observables like the expectation values of energy density $\langle T_{ab}\rangle\ell^a\ell^b$ or energy flux $-\langle T_{ab}\rangle\ell^am^b$ which are only divergent near the true singularity.

Also, from the coincidence of the behaviour from $\langle0|T_{uu}|0\rangle$ for very late times for our star and for a collapsing null shell of \cite{DFU}, our results strengthen the notion that the final regularized expectation value of the energy-momentum tensor of Hawking radiation ignores all the details of the collapse.

One could argue that, despite the dynamical similarities between 2D and 4D models with respect to the piling of outgoing rays on the horizon, the situation could be very different in four dimensions, but it must be noted that near $\mathscr I^+$, the usual conservation law $\partial_\nu T^{\mu\nu}=0$ implies \cite{Ring} that the expectation value for the 4D stress-energy tensor is proportional to $r^{-2}$ times the one in 2D for spherical waves. Assuming this to be true, conservation of energy arguments imply that the 4D values cannot explode in any non-zero measure set contained in a domain of integration, such as the event horizon.

\section{Acknowledgements}
BAC would like to thank Conselho Nacional de Desenvolvimento Cient\'ifico e Tecnol\'ogico (CNPq) for financial support under the process 200339/2014-1.



\begin{thebibliography}{1}
\bibitem{LauraB} Mersini-Houghton, L. \emph{Phys. Lett. B} \textbf{738} 61 (2014).
\bibitem{Laura} Mersini-Houghton, L. arXiv hep-th 1409.1837 (2015).
\bibitem{refB} Kawai, H; Matsuo, Y.; Yokokura, Y.; \emph{Int. J. Mod. Phys. A}, \textbf{28} 1350050 (2013).
\bibitem{Terno} Bacceti, V; Mann, R.B.; Terno, D.R.; arXiv gr-qc 1760.01180 (2017).
\bibitem{ho} Ho, P. \emph{Nucl. Phys. B}, \textbf{909} 394 (2016).
\bibitem{George} Matsas, G.E.A., Junction conditions and Stellar Collapse (unpublished).
\bibitem{rc1} Cohen, J.M.; \emph{Int. J. Theor. Phys}. \textbf{1}, 361 (1968).
\bibitem{DFU} Davies, P.; Fulling, S.; Unruh, W., \emph{Phys Rev D} \textbf{13} 2720 (1976).
\bibitem{rc2} Bellucci, S., Saharian, A.A., Yeranyan, A. H.; \emph{Phys. Rev. D} \textbf{89}, 105006 (2014)
\bibitem{daives} Davies, P.; \emph{Proc. Roy. Soc Lond. A}. \textbf{351}, 129 (1976)
\bibitem{BirrelDavies} Birrell, N.D.; Davies, P.C.; \emph{Quantum Fields in Curved Space}. Cambridge Monographs on Mathematical Physics. Cambridge Univ. Press (1982).
\bibitem{Ring} Ring, D.; \emph{Class. Quantum Grav.} \textbf{23} 5027 (2006).
\end{thebibliography}
\end{document}